\documentstyle[12pt]{article}

\def\la{\;\raisebox{-.4ex}{\rlap{$\sim$}} \raisebox{.4ex}{$<$}\;}
\def\ga{\;\raisebox{-.4ex}{\rlap{$\sim$}} \raisebox{.4ex}{$>$}\;}
\def\beq{\begin{equation}}
\def\eeq{\end{equation}}

\def\beqn{\begin{eqnarray}}
\def\eeqn{\end{eqnarray}}
\def\ra{\rightarrow}
\def\nn{\noindent}
\def\epem{e^{+}e^{-}}
\def\fb{{\rm fb}}
\def\gev{{\rm GeV}}
\relax
\begin{document}

\begin{titlepage}
\pagestyle{empty}
\rightline{CERN-TH.6932/93}
\rightline{ENSLAPP-A-425/93}
\rightline{LMU-10/93}
\vspace*{\fill}
\begin{center}
{\large{\bf Higgs Production in Association with a Vector Boson Pair \\
 at Future $\epem$ Colliders.}}
\vspace*{\fill}

\begin{tabular}[t]{c}
M.~Baillargeon$^{*,a}$, F.~Boudjema$\;^a$, F.~Cuypers$\;^b$,
E.~Gabrielli$\;^c$ and B.~Mele$\;^d$\\
\\
{$^a$ \it Laboratoire de Physique Th\'eorique, ENSLAPP $^\dagger$ }\\
{\it B.P.110, 74941 Annecy-Le-Vieux Cedex, France}\\
{$^b$ \it Sektion Physik der Universit\"at M\"unchen,}\\
{\it Theresienstra\ss e 37, D-8000 Munich 2, Germany}\\
{$^c$ \it Theory Division, CERN, CH-1211 Geneva 23, Switzerland}\\
{$^d$ \it INFN, Sezione di Roma, Italy and}\\
{ \it Dipartimento di Fisica, Universit\`a ``La Sapienza'',}\\
{ \it P.le Aldo Moro 2, I-00185 Rome, Italy}\\
\end{tabular}
\end{center}
\vspace*{\fill}

\centerline{ {\bf Abstract} }
\baselineskip=18pt
\noindent
We study Higgs boson production in association with a pair of electroweak
vector bosons ($WW,ZZ,Z\gamma$) at future $\epem$ colliders in the
framework of the Standard Model.
Total cross sections and distributions for the intermediate-mass Higgs
are presented, with special
emphasis on the Next Linear Collider (NLC) case operating at a
centre-of-mass energy $\sqrt{s} \simeq 500$ GeV,
where the cross sections turn out to be
more favourable than in larger-$\sqrt{s}$ collisions. We find that
with an integrated luminosity of 20 fb$^{-1}$
there is a sizeable event rate for the $HWW$ and $HZ\gamma$ (with a
high $p_T^{\gamma}$) channels, while a larger integrated luminosity is
needed to study the $HZZ$ production. We take into account various
backgrounds, notably top-pair and triple vector-boson production, and show
ways to significantly reduce their effects.
\vspace*{\fill}

\leftline{CERN-TH.6932/93}
\leftline{June 1993}
$\;$
\hrulefill\ $\; \; \; \; \; \; \; \; \; \; \; \; \; \; \; \; \; \; \; \;\;$
\hspace*{3.5cm}\\
{\footnotesize $\;^*$ On leave from {\em Laboratoire de Physique
Nucl\'eaire, Universit\'e de Montr\'eal, C.P. 6128, Succ. A, Montr\'eal,
Qu\'ebec, H3C 3J7, Canada.} }\\
\noindent
{\footnotesize $\;^\dagger$ URA 14-36 du CNRS, associ\'ee \`a l'E.N.S. de
Lyon et au LAPP d'Annecy-le-Vieux.}

\end{titlepage}
\baselineskip=18pt

\section{Introduction}

The clarification of the mechanism for electroweak symmetry breaking
is presently  a basic  issue for high energy physics.
One way to attack this problem is to look for Higgs bosons that arise in
the Standard Model, after spontaneously breaking  the SU(2)$\times$U(1)
electroweak symmetry.
Extensive studies have been carried out on the potential of present and
future high energy colliders for discovering Higgs bosons predicted
within and beyond the SM (for recent reviews see for instance
\cite{HHG}-\cite{saariselka}).
In this respect, $\epem$ colliders compared to
hadron machines offer the advantage of producing Higgs bosons in a particularly
clean environment. This is essential since the scalar Higgs, $H$, couples
mainly to heavy
particles, and consequently production cross sections are rather small.
Furthermore, for $m_H\la$135 GeV, $H$ decays most of the time into
the
heaviest fermion pair allowed by phase-space ($H\ra b\overline{b}$
for $m_H\ga$10 GeV).
Hence, at hadron colliders, in this range of $m_H$, Higgs detection
is made very difficult by the huge QCD backgrounds.
Present limits on $m_H$ ($m_H\ga62.5$ GeV for a Standard Model Higgs
\cite{limits})
derive from the lack of observation of a Higgs signal at LEP I.
LEP II will be able to cover the range up to $m_H \la M_Z$.
A Higgs in the intermediate mass range  80 GeV$\la m_H \la$ 140 GeV
could be observed at
future hadron colliders (LHC/SSC) only through very dedicated (and
costly) detectors \cite{2gamma}. Even in the optimistic case of
detecting a signal at LHC/SSC,
it would be impossible to study in detail the Higgs  properties,
such as its couplings, spin and parity characteristics.

On the other hand, an $\epem$ linear collider with $\sqrt{s}\simeq
(300$-$500)$ GeV
and integrated luminosity $L\simeq 10$-$20$ fb$^{-1}$ (NLC) would be an ideal
place
to observe and study in detail
an intermediate-mass or even a heavier
Higgs \cite{hamburg}-\cite{saariselka}.
At this machine, Higgs bosons would
be produced mainly through the bremsstrahlung process
$\epem\ra HZ$ and the
$WW/ZZ$ fusion processes $\epem\ra H\nu\overline{\nu}$
 and $\epem\ra H\epem$. By adding all
these contributions, one gets a cross section larger
than $100$ fb for Higgs production in the  intermediate-mass
range in the Standard Model.

In this paper we consider another class of processes that are  interesting
for Higgs studies at
future $\epem$ colliders. We consider Higgs production in association
with a pair of electroweak vector bosons
\beqn \label{HVV}
                    \epem &\rightarrow &HWW           \\
                    \epem &\rightarrow &HZZ           \\
                    \epem &\rightarrow &HZ\gamma
\eeqn
The relevant tree-level Feynman diagrams are shown in Fig. 1.
The $\gamma$ in the third channel is a high-$p_T$ observable photon.
A Standard Model Higgs is assumed.
Although of the same order in the electroweak coupling as the $WW/ZZ$
fusion processes, the new channels are suppressed because of the narrower
available phase space.
Nevertheless, we will see that production
rates for $HWW$ and $HZ\gamma$ can be  non-negligible at the NLC.

Higgs production in association with two $W$ or $Z$ can provide
further tests than those that may be probed in the fusion processes,
on the $HWW$ and $HZZ$ couplings.
Possible anomalous couplings in quadrilinear vertices, not present at
tree-level in the SM such as
$HZWW$, $H\gamma WW$, or even the C-violating $HZZZ$ or $H\gamma ZZ$
could be uniquely directly investigated in these processes
as they would cause deviations from the predicted signal.
Moreover, an accurate estimate of the channels (1)-(3) as well as some
characteristic distributions is essential, as these processes could be
potential
backgrounds for possible new physics. For instance, it has been pointed out
\cite{Bartl}
that some $ZZH$ events have the same signature as neutralino production when
the latter decays into a Higgs or a vector boson.

The plan of this paper is the following. In section 2, we describe
our procedures for computing total cross sections and
distributions for the processes under study.
We will present explicit and compact results for
the $HZ\gamma$ case and show how the matrix elements can be
factorized in terms of the two-body reactions $e^+e^- \rightarrow ZH$ times a
photonic radiator or alternatively $e^+e^- \rightarrow Z\gamma$ times a
``Higgs radiator". This Higgs radiator is also present in the other two
reactions, and for $ZZH$ production one can also write the result in a
factorized form relating it directly to $Z$ pair production.
Section 3 is devoted to the
results pertaining to total cross sections. We discuss the sensitivity
of the cross sections to the Higgs mass and the variation with
the centre-of-mass energy and compare with the other ``standard"
mechanisms of Higgs production. In section 4 we present and discuss some
characteristic distributions. In section 5 we study in detail how
these cross sections translate into numbers of events when we include
the branching fractions and when we take into account various backgrounds.
We show how the most dangerous backgrounds for $WWH$ (and $ZZH$) production
can be eliminated and how the production of three vector bosons
$WWZ, ZZZ, ZZ\gamma$ do not cause any serious problem.
Some comments together with our conclusions
are reported in section 6.

The processes in (1)-(3) have also been considered by  Barger et al.
in ref. \cite{Barger} .
We have checked that our cross sections agree with theirs.
In contrast with the previous analysis, which concentrates on
a Higgs not heavier than $50$ GeV
(already ruled out by LEP I), we cover here the case of the
intermediate-mass Higgs and perform
an extensive analysis of the  backgrounds.

\section{Description of the computation}

In this section  we describe the procedures adopted for evaluating
the cross sections and distributions for the $HVV$ processes in (1)-(3).

The Feynman diagrams  corresponding to $HVV$ processes at
tree-level are shown in Fig. 1 in the unitary gauge
(neglecting  diagrams with direct Higgs-fermion
couplings, which are suppressed by fermion masses).
They can be obtained in a straightforward way by radiating a Higgs boson
from every $W/Z$ external leg or propagator
in the set of diagrams corresponding
to the processes $\epem \ra WW, ZZ, Z\gamma$.

We have used two different and independent procedures for evaluating the
corresponding matrix elements squared,
both different from the one adopted in ref. \cite{Barger}.
In the first method we  squared the  amplitudes
summing over initial and final polarization with the help of
{\em Schoonschip} \cite{schoonship}.
The  output (which is a function
of the five independent invariants of the particular process) was then
integrated
numerically in order to get various kinematical distributions and total cross
sections. For this purpose, we used both a RGAUSS-like {\em Fortran} routine
and a {\em Vegas} Monte Carlo \cite{vegas} program to check the results.

The second way of obtaining the matrix elements squared
relied exclusively on the computer program {\em CompHep} \cite{Ilyin}.
This software generates automatically the Feynman diagrams and then yields
the matrix elements squared either in {\em Reduce} \cite{reduce} or
{\em Fortran} code.
We then fed the output of {\em CompHep} into a {\em Vegas}-based Monte Carlo
integration routine.

We do not show here the lengthy final expressions for the matrix
elements squared $|{\cal M}|^2$ for $HWW$ and $HZZ$.
Instead, we make some comments on the explicit form of $|{\cal M}|^2$ for
 $\epem \ra HZ\gamma$,
which is rather compact and exhibits some interesting features.

The matrix element squared for the reaction $\epem \ra H Z \gamma $
can be cast into the form:

\beq    \label{gamma}
|{\cal M}|^2= G^2 \frac{1}{(q^2-M_Z^2)^2}
\left( \frac{A}{p_1\cdot k} + \frac{B}{p_2\cdot k} + \frac{C}{(p_1\cdot k)
(p_2\cdot k)}
\right),
\eeq

\noindent where the momenta are defined as $e^-(p_1) e^+(p_2) \ra
H(h) Z(z) \gamma(k)$, with $q=h+z=p_1+p_2-k$ and $P=p_1+p_2$
$\;\;(P^2=2 p_1\cdot p_2=s)$ and with

\beqn
A&=&p_2\cdot k + \frac{2}{M_Z^2} \;(p_2\cdot z)(k\cdot z) \;\;\;\;\;\;
\;\;\;\ B=A(p_2 \ra p_1)    \nonumber \\
C&=& s \left(\frac{s}{2}-k\cdot P\right) +   \nonumber \\
 \frac{2}{M_Z^2}&\cdot&\left( (p_1\cdot z)(p_2\cdot z) (s-k\cdot P) \;
-\frac{s}{2} \; (P\cdot z)(k\cdot z)
+ (p_1\cdot  z)^2(p_2\cdot k) +
    (p_2\cdot z)^2(p_1\cdot k) \right)  \nonumber
\eeqn
\noindent
The constant $G$ in eq. (\ref{gamma}) is defined as
\beq
G^2=4 \pi^3 \alpha^3 \frac{M_Z^2}{s_W^4 c_W^4} (1+(1-4 s_W^2)^2).
\eeq

Equation (\ref{gamma}) clearly displays
the collinear and the soft-photon singularities. These are regularized
by imposing a $p_T$ cut on the photon such that
$p_T^{\gamma}>p_T^{cut}$, which leads to a description of the $ZH\gamma$
production
with an observable photon in the final state.

It is interesting to note that eq. (\ref{gamma}) shows factorization
in the limit of both soft photons and soft Higgses. The notion of a
``soft Higgs" refers to a situation where the Higgs is massless and has
a very small energy.
Let us analyse in detail the two cases.

\subsection{Factorization of the low-energy photons}

We can readily recover the low-frequency collinear photons by only
keeping the terms that are most singular in $1/|k_0|$ in eq. (\ref{gamma}).
These are obtained by letting $k \ra 0$ in $A,B$ and $C$, which shows that only
the cross-term $C$ survives. Therefore, the contributing terms to the
leading-log approximation are:

\beq
|{\cal M}|^2 \ra  G^2 \frac{1}{(q^2-M_Z^2)^2} \frac{s^2}{2}
\frac{1}{(p_1\cdot k)(p_2\cdot k)}
\left(1+  \frac{4}{s M_Z^2}\; (p_1\cdot z)(p_2\cdot z) \right),
\eeq
\noindent
which agrees with the low-energy factorization:

\beqn
|{\cal M} (e^+ e^- \ra H Z \gamma)|^2 &\ra&
|{\cal M}(e^+ e^- \ra H Z )|^2 \left( - e^2 g_{\alpha \beta} (
\frac{p_1^{\alpha}}{p_1\cdot k}-\frac{p_2^{\alpha}}{p_2\cdot k} )(
\frac{p_1^{\beta}}{p_1\cdot k}-\frac{p_2^{\beta}}{p_2\cdot k} ) \right)
\nonumber \\
&=& e^2 \frac{s}{(p_1\cdot k)(p_2\cdot k)} |{\cal M}(e^+ e^- \ra H Z )|^2
\eeqn

Indeed we recognize the first term in the last expression to be
the ``photon radiator" and we check, by explicit calculation, that the two-body
process $e^+ e^- \ra HZ$ is given by

\beqn
|{\cal M} (e^+ e^- \ra H Z )|^2 =
(G/e)^2 \frac{1}{(q^2-M_Z^2)^2} \frac{s}{2}
\left(1+  \frac{4}{s M_Z^2} \; (p_1\cdot z)(p_2\cdot z) \right).
\eeqn

Note that, at the lowest order, the total integrated cross section for
$ZH$ production writes:

\beq
\sigma(HZ)=\frac{\pi \alpha^2}{48 s_W^4 c_W^4} (1+(1-4 s_W^2)^2)
\frac{M_Z^2}{(s-M_Z^2)^2} \; \beta \left( 3 + \frac{s \beta^2}{4 M_Z^2}
\right),
\eeq

\noindent
where we have defined

\beq
s \beta = \sqrt{ (s-(M_Z+m_H)^2)(s-(M_Z-m_H)^2)}.
\eeq

For later reference, let us point out that for large momentum Higgs,
the $Z$ are produced with a predominantly longitudinal polarization and
are hence essentially orthogonal to the beam. We recall that the angular
distribution is written as

\beqn
\frac{{\rm d}\sigma}{{\rm d}\cos \theta} \propto (1+\cos^2 \theta)
+ \frac{s}{4 M_Z^2} \left(
1+\frac{M_Z^2}{s}-\frac{m_H^2}{s}\right)^2 \sin^2 \theta .
\eeqn

The first term, $(1+\cos^2 \theta)$, represents the transverse $Z$
contribution,
while the longitudinal one has an ``enhanced" coupling: $s/M_Z^2$.

\subsection{Factorization in low-energy, low-momentum Higgses}

 One can also factorize out the ``Higgs radiator" by going into
the massless low-energy Higgs limit, although the present limit on the mass
of the Higgs makes this a purely but interesting ``academic exercise" (which
can nonetheless be used as yet an additional check on our computations). The
cross section in this case can be written as the product of the
$e^+ e^- \ra Z \gamma$ cross section
times the ``Higgs radiator".
One first verifies that the coefficients of $1/M_Z^2$ in $A, B$ and $C$
vanish in the limit $h \ra 0$, and that the remaining parts of $A, B$ and $C$
add up to give the $e^+ e^- \ra Z \gamma$ matrix element squared times
a Higgs radiator, that is the propagator factor $1/(q^2-M_Z^2)^2$
modulo the $HZZ$ coupling $g_{HZZ}^2$ :

\beq
|{\cal M} (e^+ e^- \ra H Z \gamma)|^2 \ra
\frac{g_{HZZ}^2}{(q^2-M_Z^2)^2}  |{\cal M}(e^+ e^- \ra Z \gamma )|^2 .
\eeq

In fact this factorization works also in the case of $HZZ$ production.
Denoting the external $Z$ momenta by $z_1$ and $z_2$, the amplitudes may be
written in the limit of a soft Higgs:

\beq
{\cal M} (e^+ e^- \ra H Z Z) \ra
\left(\frac{g_{HZZ}}{(z_1+h)^2-M_Z^2}  +
\frac{g_{HZZ}}{(z_2+h)^2-M_Z^2} \right) {\cal M}(e^+ e^- \ra Z Z).
\eeq

In the case of $W^+W^-H$ production, this factorization fails due to the
emission by
the ``internal" $Z$ line. However we would like to point out, based on the
analogy with the ``backbone" reaction $e^+e^- \ra W^+ W^-$, that at threshold
one expects a dominance of the neutrino-exchange diagrams. Indeed the $P$-wave
nature of the $s$-channel $Z$ and $\gamma$ exchanges means that these are
suppressed at threshold. We have checked this numerically. For instance,
for a Higgs mass of $90$ GeV, the approximate ($t$-channel) and total
cross sections compare as follows

\vspace*{0.5cm}
\begin{center}
\begin{tabular}{|c|c|c|c|}
\hline
$\sqrt{s} {\rm (GeV)}$&t-channel only (fb)&Total (fb)&Relative error
\\ \hline
260&0.38&0.37&-3$\%$ \\ \hline
270&1.3&1.2&-8$\%$ \\ \hline
300&5.7&4.2&-26$\%$ \\ \hline
\end{tabular}
\end{center}
\vspace*{0.5cm}

\noindent
We see that within $10$ GeV about the threshold, the agreement is better
than $3\%$; however, already at $300$ GeV the $s$-channel is badly required.

\section{Total cross sections}

In this section we study total cross sections for $HVV$ production
as a function of the Higgs mass $m_H$ and the $\epem$ centre-of-mass energy
$\sqrt{s}$.
We concentrate on an intermediate-mass Higgs and 0.3 TeV$\la \sqrt{s}\la 2$
TeV.
The cross sections are calculated with $M_Z=91.18$ GeV, $M_W=80.1$ GeV and we
take an effective $\sin^2 \theta_W=0.232$. Moreover, apart from the case of
$HZ\gamma$ where we take the electromagnetic coupling constant for the
real photon at $q^2=0$ (i.e $\alpha(0)=1/137$), we use
 $\alpha(M_Z^2)\simeq1/128$.

Since, as we shall see below, at centre-of-mass energies around
$\sqrt{s}\simeq$ 500 GeV the production rates for $\epem\ra HWW,HZZ$
in the case of an intermediate-mass Higgs are the largest, we show in fig. 2
a comparison, at $\sqrt{s}=$ 500 GeV, of our three processes with the
main production channels of the Higgs,
that is  the bremsstrahlung process $\epem\ra HZ$ and the
$WW/ZZ$ fusion processes $\epem\ra H\nu\overline{\nu}$
 and $\epem\ra H\epem$.
The cuts $p_T^{\gamma}>10$ GeV and $|y^{\gamma}|<2$ are imposed on
the $HZ\gamma$ process. We can observe that for intermediate $m_H$
the rates for $HWW$ are comparable to those for $H$ production
via $ZZ$ fusion. In particular, assuming an integrated luminosity of
20 fb$^{-1}$, one gets a sizeable (200-60) $HWW$ events (not including
branching fractions) by varying $m_H$
in the range $(M_W-2M_W)$. The $HZ\gamma$ channel is a bit lower
than $HWW$ for intermediate $m_H$, but exceeds it for larger Higgs masses.
The $HZZ$ process has the lowest rate and will need higher integrated
luminosity in order to be studied also in the intermediate-Higgs range.
Although Higgs production through both the $WW$ fusion process and the
Bjorken process is about an order of magnitude higher than through $HWW$ or
$HZ\gamma$ production, the latter reactions are a welcome additional means
of producing sizeable numbers of Higgs events.
Even for Higgs masses up to $m_H\simeq$ (250-300) GeV one still expects
a few $HWW$ raw events per year to be produced  at the NLC.
In the $HZZ$ case, a few events are still collected
up to $m_H\simeq$ (150-200) GeV.

\begin{table}
\begin{center}
\begin{tabular}[tbh]{|c|c|c|c|} \hline
$m_H$ (GeV)&$\sigma(HWW)$(fb)&$\sigma(HZZ)$(fb)&$\sigma(HZ\gamma)$(fb) \\
\hline
 60 &\ 15.2   &\  1.43  &\ 5.08  \\ \hline
 70 &\ 12.7   &\  1.21  &\ 4.66  \\ \hline
 80 &\ 10.7   &\  1.03  &\ 4.33   \\ \hline
 90 &\ 9.05   &\  0.874  &\ 4.02   \\ \hline
 100 &\ 7.71   &\ 0.749   &\ 3.74   \\ \hline
 110 &\ 6.59   &\ 0.643   &\ 3.49   \\ \hline
 120 &\ 5.63   &\ 0.551   &\ 3.25   \\ \hline
 130 &\ 4.82   &\ 0.472   &\ 3.03   \\ \hline
 140 &\ 4.12   &\ 0.403   &\ 2.82   \\ \hline
 150 &\ 3.51   &\ 0.343   &\ 2.62   \\ \hline
 160 &\ 2.98   &\ 0.291   &\ 2.42   \\ \hline
 170 &\ 2.52   &\ 0.245   &\ 2.24   \\ \hline
 180 &\ 2.12   &\ 0.204   &\ 2.07   \\ \hline
 190 &\ 1.77   &\ 0.169   &\ 1.90   \\ \hline
 200 &\ 1.47   &\ 0.138   &\ 1.74   \\ \hline
 210 &\ 1.21   &\ 0.112   &\ 1.59   \\ \hline
 220 &\ 0.978   &\ 0.0888  &\ 1.44   \\ \hline
 230 &\ 0.783   &\ $-$   &\ 1.30   \\ \hline
 240 &\ 0.617   &\ $-$   &\ 1.17   \\ \hline
 250 &\ 0.476   &\ $-$   &\ 1.05   \\ \hline
 260 &\ 0.359   &\ $-$   &\ 0.929   \\ \hline
 270 &\ 0.262   &\ $-$   &\ 0.818   \\ \hline
 280 &\ 0.183   &\ $-$   &\ 0.713   \\ \hline
 290 &\ 0.121   &\ $-$   &\ 0.616  \\ \hline
 300 &\ 0.0737   &\ $-$   &\ 0.525   \\ \hline
 \end{tabular}
\vskip .3cm
{\bf Table I}: Total cross sections for $\epem \ra HWW,HZZ,HZ\gamma$
at $\sqrt{s}=500$ GeV vs. $m_H$.
The cuts $p_T^{\gamma}>10$ GeV and $|y^{\gamma}|<2$ are applied to
the final photon in the process $\epem \ra HZ\gamma$.
\end{center}
\end{table}
\noindent
In Table I we give again total cross sections
at $\sqrt{s}=500$ GeV vs. $m_H$.
Cross sections for the $HWW$ and $HZZ$ processes are also
shown in figs. 3 and 4.
In fig. 3 the $HWW$ cross section is plotted  versus $\sqrt{s}$
for  intermediate $m_H$.
We can observe that the $HWW$ production is peaked for values of the
centre-of-mass energy around
$\sqrt{s}\sim 500$ GeV for $90<m_H<150$ GeV. In particular,
$\sigma_{MAX}\simeq 9$ fb for
$m_H=90$ GeV, while at $\sqrt{s}\simeq 2$ TeV,  production rates are
about four times smaller for the same $m_H$. The same pattern
holds for the $HZZ$ production (cf. fig. 4), but the $HZZ$ yield is
about ten times smaller than that of $HWW$, for same
$m_H$ and $\sqrt{s}$. This is mainly because the $Z$ has weaker
couplings to the initial fermions than the $W$.

Cross sections for the $HZ\gamma$ channel are shown in figs. 5 and 6.
In order to select high-$p_T$ observable photons we impose a cut
of 10 GeV or more on the $\gamma$ transverse momentum.
We also cut on the photon pseudorapidity imposing everywhere
$|y^{\gamma}|<2$.

In fig. 5 we plot the total cross section versus
 $\sqrt{s}$ in the range $0.2<\sqrt{s}<3$ TeV, for $m_H=90,120,150$ GeV
 and $p_T^{\gamma}>10$ GeV.
Compared to $HWW$ and $HZZ$, cross sections for $HZ\gamma$ are
 peaked in a lower range of $\sqrt{s}$, due to the altogether
lighter final state. For example, for $m_H=$90 GeV
$\sigma(HZ\gamma)$ has its peak at $\sqrt{s}\simeq 300$ GeV,
where $\sigma(HZ\gamma)\simeq 10$ fb.

Figure 6  shows $\sigma(HZ\gamma)$
versus $m_H$ at $\sqrt{s}=300$ GeV and 500 GeV.
The effect of increasing the cut on $p_T^{\gamma}$ is also shown in both
cases. We can see that for $p_T^{\gamma}>10$ GeV, a lower-$\sqrt{s}$
machine can be better than the NLC with $\sqrt{s}\simeq 500$ GeV for
studying $HZ\gamma$ with $m_H\la 150$ GeV.
This is no longer true if one increases the cut on $p_T^{\gamma}$ to
about 40 GeV or more, since for small cuts
on $p_T^{\gamma}$, the behaviour of $\sigma(HZ\gamma)$ with $\sqrt{s}$
is closer to the one of $\sigma(HZ)$, which indeed decreases as $1/s$.

\section{Characteristic distributions}

\subsection{$\epem \ra H Z \gamma$}
The various distributions in this process can be easily understood
when viewing the photon  radiated from the electron/positron
legs as a photon of a predominantly bremsstrahlung nature:
the cross section is
largest for the lowest-$p_T$ photons independently of the mass of the Higgs.
This is well rendered in fig. 7.1. This feature has an impact on the
characteristics of the $Z$ and $H$ distributions, which are to be likened
to those in the process $\epem \ra ZH$. In fact the energies of both the
$Z$ and the $H$ tend to peak around the beam energy, especially for small
values
of the $p_T^\gamma$ cut as exemplified in figs. 7.2 and 7.4,
where $p_T^\gamma>10$ GeV. For
higher values of this cut these spectra are slightly broader and away from
$\sqrt{s}/2$ due to the reduction in the ``effective" centre-of-mass energy
of the process $\epem \ra ZH$ as shown in figs. 7.3 and 7.5 for
$p_T^{\gamma}=30$ GeV.
Anyhow, in all
cases, the Higgs is emitted preferentially with a high $p_T$
(see figs. 7.6 and 7.7). In fact the Higgs and the $Z$ (since the photon
has a low energy and prefers to be collinear to the beam most of the time)
favour the central region, especially for low $p_T^\gamma$ and
smaller Higgs masses (figs. 7.8-7.11). This is again supported by the fact
that the $Z$ and the Higgs are almost back-to-back (see figs. 7.12 and 7.13).
This feature of the Higgs takes its
source from the process $\epem \ra ZH$ and reflects
the fact that the $Z$, at high energies, is almost longitudinal. This
particular
distribution, as we will see below, is common to all three reactions that
we studied.
Probably more telling is the scatter plot, which confirms that
the events cluster around
$E_H\simeq E_{beam}$ while, at the same time, $E_\gamma$ is essentially
below $30$ GeV (fig. 7.14). For $m_H=120,150$ GeV  ($p_T^\gamma > 10 GeV$),
the scatter plot is very similar to the one in fig. 7.14.
These characteristics do not change
significantly when we move to a
higher centre-of-mass energy as depicted, for $\sqrt{s}=1$ TeV in
figs. 7.15 and 7.16.

\subsection{$\epem \ra HZZ $}
At $500$ GeV the shape of the various distributions are not sensibly different
for our three
choices of the intermediate-Higgs mass, $m_H=90,120,150$ GeV, apart from
the distribution
in the energy of the Higgs, $E_H$. First, the transverse momentum of the
Higgs  (see fig. 8.1)
is broadly
peaked around $100$ GeV, which is about half the value allowed by the
kinematics.
Once again the Higgs is produced centrally (fig. 8.2). It is
preferentially within about $30^{\circ}$ from the plane
orthogonal to the beam direction. These two facts explain the shape of
the distribution in the energy of the Higgs. Given
$p_T^H$, we essentially have $E_H\sim \sqrt{(p_T^H)^2 + m_H^2}$. With respect
to the most energetic $Z$, the Higgs is emitted preferentially in the opposite
direction, in fact the two particles are almost back to back as shown in
fig. 8.3
where we see that this spectrum has a smooth
hump around a value $\sim 160^{\circ}$. On the other hand,
with respect to the least energetic $Z$,
the Higgs is rather orthogonal (fig. 8.4). A very useful
and revealing distribution is exhibited as a scatter plot in fig. 8.5
(for $m_H=120,150$ GeV, the plot is very similar to the one for $m_H=90$ GeV).
This plot
shows that the most favourable situation is when both the Higgs and the
most energetic $Z$ are orthogonal to the beam direction. These features
are all to be compared with the previous reaction, $\epem \ra HZ\gamma$.
One can liken the r$\hat{o}$le of the least energetic $Z$ to that played
by the photon,
the $Z$ mass providing in a sense a natural ``energy" cut. Then, while we
expect the least energetic $Z$ to be emitted off the electron line, the most
energetic $Z$ is preferentially produced as a longitudinal $Z$
together with
the Higgs, i.e., it is radiated by the $Z^\star ZH$ vertex. Of course,
since the polarization vector of a longitudinal energetic $Z$ introduces the
enhancement factor $E_Z/M_Z$, and the $Z$ current in $Z^\star ZH$ is not
conserved (or rather not transverse), we can understand the fact that the $ZZH$
cross section is largest when it is the most energetic $Z$ that is emitted
from the $ZZH$ vertex. Since this is a situation akin to the Bjorken process,
one can also understand that the energetic $Z$ favours the central region.

\subsection{$\epem \ra HW^+W^- $}
A comparison among the shapes of the distributions in the transverse
momentum, the angle with respect to the beam, and the
energy of the Higgs in the reaction
$\epem \ra H W^+W^- $ and $\epem \ra H ZZ$, shows that these are almost
identical
for all three representative values of the Higgs mass
$m_H=90,120,150$ GeV (figs. 9.1-9.3).
This a reflection that, modulo the strengths of
the couplings of the $W$ and the $Z$ to the electron and Higgs, these
distributions are dictated by the behaviour of the $t$-channel. For $WWH$
this is a behaviour similar to the one exhibited by the $t$-channel
neutrino-exchange diagram in
$e^+e^- \ra W^+ W^-$ \footnote{The shape of the angular
distribution in  $W^+W^-$ production reflects essentially the $t$-channel
exchange.}.
After all, $\epem \ra W^+W^-H $ is directly  related to
$WW$ by ``grafting" a Higgs.
This observation is further
supported by the fact that $W^-$ favours being emitted in the forward
direction, the electron direction, as fig. 9.4 shows. In this case, as the
scatter plot in fig. 9.5 shows, the $W^-$ tends to
take the maximum kinematically allowed energy.

\nn Figure 9.6  shows
that the Higgs prefers to be in the direction perpendicular to the beam, while
at the same time the $W^-$ is almost exclusively
in the forward hemisphere,
$\theta_{e^- W^-} < 90^{\circ}$. The $W^-$ distribution around the electron
direction
is a reflection of the fact that half of the time it is emitted longitudinally
off the Higgs (hence it would rather be perpendicular to the beam) and
half of the time (when the Higgs is emitted by the $W^+$) the $W^-$ is
produced in the very forward region, as in
$\epem \ra W^+W^-$.

\nn These
features remain essentially unaltered for the three values of $m_H$ that we
considered.

\section{Signatures and backgrounds}
The intermediate-mass Higgs that we consider will decay predominantly into
$b\bar b$. All the processes we have studied will therefore
consist of at least one
pair of $b$ quarks with
a high transverse momentum as we saw in the
previous section; $b$ tagging with a good vertex detector will be very
helpful. We note that the $b$ pair should be reconstructed with an invariant
mass around\footnote{Allowing for the experimental resolution.}
the Higgs mass, which should be well measured in the fusion or/and
the Bjorken process. The distribution of the $b$ quark is isotropic in the rest
frame of the Higgs, contrary, for instance, to $b$ quarks emanating from
a $Z$, which are distributed mainly\footnote{Standard $Z$'s at these
energies are essentially transverse in processes not
involving the Higgs.} according to $(1+\cos^2 \theta^{\ast})$,
where
$\theta^\ast$ is the angle measured in the $Z$ rest frame between the
decaying $b$ and the axis corresponding to the $Z$ flight direction.
Therefore, in principle, reconstruction of these
distributions could help in Higgs detection. However, one needs a large enough
sample of $b$ in order to reconstruct these angular distributions,
so that for the reactions we consider it will suffice to tag
the $b$.
One type of potential background (especially for $m_H \sim M_Z$)
to the reactions we study are precisely those
where an $H$ is replaced by a $Z$, namely $e^+ e^- \ra WWZ,ZZZ,ZZ\gamma$, with
one of the $Z$ decaying subsequently into $b\bar b$. At centre of mass of
500 GeV, the cross sections for these three reactions are \cite{quarticplb}

\beqn
\sigma(e^+ e^- \ra W W Z)&=& 39 ~\fb \;\; {\rm with} \;\; m_H < 2M_W \nonumber
\\
\sigma(e^+ e^- \ra ZZZ)&=& 1 ~\fb {\rm \;\;with} \;\; m_H < 2M_W \nonumber \\
\sigma(e^+ e^- \ra ZZ\gamma)&=& 15 ~\fb {\rm \;\;for} \;\; p_t^{\gamma}>20~\gev
 \;,\; |y^\gamma|<2
\eeqn

\noindent
As the $Z$ branching ratio into $b$'s is $\sim 15\%$,
$b$ tagging will help considerably since otherwise one has to consider a
$Z$ into jets branching ratio of about $70\%$. Hence, $b$ tagging reduces
these eventual $Z$-initiated
backgrounds by more than a factor 4. By taking any of the
$Z$ into $b \bar b$, while the other weak bosons can decay into
anything, one gets

\begin{eqnarray}
\sigma(e^+ e^- \ra W W Z \ra W W b \bar b)&=&
5.85 ~\fb {\rm \;\;with} \;\; m_H < 2M_W \nonumber \\
\sigma(e^+ e^- \ra ZZZ \ra Z Z b \bar b)&=& 0.45~\fb {\rm \;\;with} \;\;
m_H < 2M_W
\nonumber \\
\sigma(e^+ e^- \ra ZZ\gamma\ra Z \gamma b \bar b)&=&4.5 ~\fb {\rm \;\;for} \;\;
p_t^{\gamma}>20~\gev \;,\;
|y^\gamma|<2
\end{eqnarray}

\noindent
These backgrounds are most serious for $m_H \sim M_Z$. However, in this case,
the $H$ branching ratio into $b \bar b$ is about $85\%$.
At $\sqrt{s}=500$ GeV, we find for $m_H=M_Z$,
$\sigma(WWH\ra WW b \bar b)\simeq 9 \times 85\%\simeq 7.7$ fb and
$\sigma(ZZH\ra ZZ b \bar b)\simeq 0.87 \times 85\% \simeq 0.74$ fb
which are sensibly
above the three-vector-boson background. On the other hand,
$\sigma(Z\gamma H \ra Z \gamma b \bar b) \simeq 3.2 \times 85\% \simeq 2.7$ fb
for $p_t^{\gamma}>20$ GeV, which is  below, but comparable, to
the corresponding three-vector-boson background.
Therefore the Higgs signal can easily be disentangled from triple vector-boson
production. For larger Higgs masses, both the Higgs
cross sections and $Br(H\ra b \bar b)$ drop.
For instance, for $m_H=120$ GeV, one gets
$\sigma(WWH\ra WW b \bar b)\simeq 5.6 \times 70\% \simeq 3.9$ fb,
$\sigma(ZZH\ra ZZ b \bar b)\simeq 0.55 \times 70\%=0.39$ fb and
$\sigma(Z \gamma H\ra Z \gamma  b \bar b)\simeq 2.5 \times 70\%=1.7$ fb.
If $m_H=150$ GeV, one collects ``only"
$\sigma(WWH\ra WW b \bar b)\simeq 3.5 \times 50\% \simeq1.75$ fb,
$\sigma(ZZH\ra ZZ b \bar b)\simeq 0.34 \times 50\%=0.17$ fb and
$\sigma(Z \gamma H\ra Z \gamma  b \bar b)\simeq 1.9 \times 50\%=1$ fb.
However, although the signal for $m_H=150$ GeV is about three to four times
smaller than the corresponding three vector-boson background, the invariant
mass of the $b\bar b$ system is such that it should not be mistaken as coming
from the $Z$, hence almost eliminating this background.
Therefore, the three vector-boson production does not seem to pose any
serious problem for $VVH$ detection.

\nn In fact a ``huge" background to $WWH$ detection comes
from top-pair production with the top decaying exclusively into $Wb$,
leading to a topology $W^+ W^- b \bar b$. The cross section for
this process
at $\sqrt{s}=500$ GeV is $\sim 660$ fb for $m_t=150$ GeV. This is about
two orders of magnitude above the signal. To study and show ways to
reduce this background, we will take the representative value of $m_t=150$ GeV.
The following discussion does not change much for other values of $m_t$
favoured by LEP I data. Even in this situation, $b$ tagging is crucial.
In order of reducing the $t \bar t$ background,
one can impose a cut on the invariant mass of the
$b\bar b$ system, $m_{b\bar b}$, within $10$ GeV of the Higgs mass.
 A {\em Pythia}-based simulation of
$t \bar t$ events, with
subsequent decays of top into $b$
\footnote{The full spin-correlations are kept. We are grateful to G. Azuelos
for running {\em Pythia} for us.}, reveals that the $m_{b\bar b}$ distribution
shows a broad hump  around values corresponding to an intermediate $m_H$.
For both $M_Z - 10$ GeV $< m_{b\bar b} < M_Z + 10$ GeV and
110 GeV $< m_{b\bar b} < 130$ GeV (relevant for $m_H=120$ GeV)
we find that there are
still $10\%$ of the $t \bar t$ events  that pass this cut.
 This still amounts to
a cross section of $\sim 62$ fb. We found that a much more efficient and
simple selection criterion was to reject all the Higgs events in $WWH$ that
simulate $t \bar t$ when the invariant mass of both the tri-jet (in our
case $Wb$) system falls within $\pm 15$ GeV of the top mass.
Once again, to
reduce as much as possible the error in assigning the jet to its parent
particle, $b$-tagging will be extremely useful. This is because, out of the
six jets, the $W$ is experimentally reconstructed by only ``pairing" the
4 non-$b$ jets,
so that each pair recombines to give the $W$ mass. We would then
recombine the $Wb$ system to give the top.
As we want to exploit a good vertex detector for $b$-tagging without
charge identification\footnote{This is because the charge analysis is
necessarily done with a much reduced, $\sim 10\%$, sample of $b$ relying on
the high-$p_T$ lepton
from the semileptonic decay of the $b$ and would entail a considerable
loss in our signal of Higgs.},
and since by using the hadronic
decays of both $W$ it would be extremely difficult
to reconstruct their charges anyway, we tried
 in our program both combinations of $Wb$ to reconstruct the top.
To perform this analysis, we included the $H$ decay  into $b\bar b$
by first taking an isotropic distribution in the Higgs rest frame then
boosting the events in the laboratory frame. We then demand that all
$WWH\ra WWb ``b'"$ events passing the {\em simultaneous} cuts

\beqn
m_t - 15~\gev < M_{W^+b} < m_t + 15~\gev \;&{\rm and}&\;
m_t - 15~\gev < M_{W^{-}b'} < m_t + 15~\gev \nonumber \\
&{\rm or}& \nonumber \\
m_t - 15~\gev < M_{W^+b'} < m_t + 15~\gev \;&{\rm and}&\;
m_t - 15~\gev < M_{W^-b} < m_t + 15~\gev \nonumber \\
\hspace*{2cm}
\eeqn

\noindent
{\em not} be counted as a $WWH$ signal. We find that the number of events is
practically unaltered by this cut (the loss is about $3\%$):

\vspace*{0.5cm}
\begin{tabular}{|c|c|c|c|}
\hline
$m_H(GeV)$&$\sigma(WWH)$ (fb) before cut & $\sigma(WWH)$ (fb)
after cut& Percentage loss \\ \hline
90&9.05&8.77&3.1$\%$ \\ \hline
120&5.63&5.45&3.2$\%$ \\ \hline
150&3.51&3.37&4.$\%$ \\ \hline
\end{tabular}
\vspace*{0.5cm}

We should add that this method should also work when one of the $W$ decays
leptonically, as there are enough constraints to reconstruct the neutrinos
and hence both invariant masses. Let us point out that a reconstruction
of $t \bar t$ events away from threshold, based on the 3-jet clustering, was
conducted in \cite{top500}. It shows that a tail remains. But this tail is
mainly due to misassigned jets. This combinatorial error, as we noted above,
will be much reduced if a preliminary identification of the two $b$ jets is
 done. In practice, this tail will also be further reduced by imposing
our first cut on the invariant mass of the $b \bar b$ system, which cuts the
$t \bar t$ by an order of magnitude. We conclude that $t \bar t$ is not a
problem.

With at least one $W$ decaying into jets, and not taking into account decays
into $\tau$'s, the useful combined branching fraction of the $WW$ is as large
as $77\%$. After the cut on the $t \bar t$ ``misidentification",
one gets a clean number of events of reconstructible $WWH$ at
$\sqrt{s}=500$ GeV. Assuming
an integrated luminosity of ${\cal L}=20$ fb$^{-1}$, one collects:

\beqn
N_H \sim 115 \;\;\;\;\;\; {\rm for}\;\;\;\; m_H=90~\gev \nonumber \\
\;\;N_H \sim 60 \;\;\;\;\;\;\;\; {\rm for} \;\;\;\;m_H=120~\gev  \\
\;\;N_H \sim 26 \;\;\;\;\;\;\;\; {\rm for} \;\;\;\;m_H=150~\gev  \nonumber
\eeqn

We see that we have a healthy number of events and even if we take an overall
efficiency of $50\%$, it is still observable for $m_H$ as large as
$150$ GeV.

For the $ZZH$ signal with $m_H=90$ GeV one has $\sigma(ZZH\ra ZZ b \bar
b)\simeq
0.74$ fb, which is above the corresponding $3Z$ background. Allowing both $Z$
to decay into \underline{non $b$} jets,
once again we will have a large background due to
top-pair production, with both $W$'s decaying hadronically, as it will be
difficult to disentangle a dijet invariant mass clustering around $M_W$ from
one clustering around $M_Z$.
In principle, we should apply the same 3-jet veto as for the $WWH$ production
to cut the $t \bar t$ background. However, the signature (when
the jet from the $Z$ is not ``tagged" as a $b$ quark) is the same
as for $WWH$, with the $W$ decaying into hadrons.
Therefore we suggest that, for this particular signature, we should just
add the $ZZH$ events to those from $WWH$ since these few $ZZH$
events represent about a tenth of the similar $WWH$ events.
We do not attempt to find criteria to disentangle these $ZZH$ events from
the $WWH$ ones because, for both processes, the
distributions in the variables of the weak vector bosons and the Higgs are
very similar (see section 4).
Moreover, even for $m_H \sim 90$ GeV, taking only the hadronic
decays  not containing  $b$ quarks for both $Z$ only amounts to about
$5$ events with ${\cal L}=20$ fb$^{-1}$.
Events that would not be mistaken as coming from $WWH$
arise from only one $Z$
decaying into jets while the other decays into leptons (in this case
mostly neutrinos, i.e. large transverse missing energy). This corresponds to
a combined branching fraction
$B_{comb.} \sim B(H\ra b \bar b)\times 0.36$
\footnote{In all cases we have not included $Z\ra \tau \bar{\tau}$.}.
Alternatively one could demand
that one $Z$ decays into $b$'s while the other is allowed
to decay into anything corresponding to
$B_{comb.} \sim B(H\ra b \bar b)\times 0.23$
\footnote{We have not double counted the $l \bar l b \bar b$, present in the
previous sample.}. In this situation we will
have four $b$ jets. Taking
into account both signatures we end up with an almost background-free
branching ratio of $B_{comb.}\simeq B(H\ra b \bar b)\times 0.59$.
Assuming an integrated luminosity of $20$ fb$^{-1}$
one would collect $\sim 10$ Higgs events, through $ZZH$
production for a Higgs mass of $90$ GeV and about 6  for a mass of $120$ GeV
at $\sqrt{s}=500$ GeV. At 1TeV taking a luminosity of $60$ fb$^{-1}$ these
rates
are 14 and 11 events respectively.

The $HZ\gamma$ production does not suffer from the top pair production
background. One should however still insist on $b$ tagging. In fact,
due to the large $Z$ branching ratio into jets,
the bulk of the $HZ\gamma$ events will consist of four jets and a photon.
This is the same signature as the radiative $W$ pair production process, i.e.,
$e^+ e^- \ra WW\gamma$ with both $W$ decaying into jets.
At $\sqrt{s}=500$ GeV and with $p_T^\gamma >20$ GeV and $|y^\gamma|<2$, this
process has a cross section of $66$ fb
\cite{quarticplb} (after folding with the branching ratios for $W \ra jj$).
This background is more important when $m_H \sim M_Z$, but  again
it should be under control after tagging the $b$-jets from $H$.
The only potential background left when $b$-tagging is effective is due to
$ZZ\gamma$ production, with one $Z$ decaying into $b$ quarks, especially
when $m_H \sim M_Z$.  Imposing the cuts $p_T^\gamma >20$ GeV and
$|y^\gamma|<2$,
and allowing the second $Z$ from $ZZ\gamma$ to decay into anything (in the
case of $\nu \bar \nu$, one will require a large missing $p_T$),
 one gets about 90 events
at $\sqrt{s}=500$ GeV with an integrated luminosity of $20$ fb$^{-1}$.
 This is to be contrasted with the $HZ\gamma$ signal which,
for a $Z$ decaying into anything will produce about 53 events for
$m_H=90$ GeV, 34 for $m_H=120$ GeV and  18 for $m_H=150$ GeV. Hence,
for $m_H\sim M_Z$, when this background is more dangerous,
the signal clearly stands out.
For this value of the Higgs mass, the ratio of signal over background is
$S/B \simeq 0.6$.

\section{Conclusions}
 One of the primary motivations for the construction of a
 linear $e^+e^-$ collider with $\sqrt{s} \sim 500$ GeV is the production
and the study of the properties
of the Higgs with an intermediate mass ($M_W\la m_H \la 2M_W$).
Such a Higgs will be difficult to
detect at the planned $pp$ machines. In this paper we have investigated
new mechanisms for the production of the Higgs in $e^+e^-$ collisions,
namely the
associated production of the Higgs with a pair of vector bosons, taking
advantage of the large $WW$ and $Z\gamma$ cross sections to which
we have ``grafted" a Higgs.

\nn We find that although Higgs production through
$WWH$ and $ZH\gamma$ are about an order of magnitude smaller than for the
main Higgs production mechanisms
through $WW$ fusion or $ZH$ production, the number of events
one collects with an integrated luminosity of ${\cal L}=20$ fb$^{-1}$ at a
centre-of-mass energy of $500$ GeV is quite substantial. We have shown how
some processes,
which can at first be considered as serious backgrounds (like top pair
production and triple vector-boson productions), can be efficiently eliminated,
especially by requiring $b$-tagging. Leaving aside the issue of
detection efficiencies and systematics, which can only be reliably estimated
with a proper detector simulation, but taking into account the observable
decays of the final particles, we find that one can have about $120$
$WWH$ events for $m_H \simeq 90$ GeV and about $60$ for a Higgs mass of
$120$ GeV,  not including those events which may
simulate some backgrounds.
The $ZH\gamma$ process also provides another $50$ ($m_H=90$ GeV)
or $30$ ($m_H=120$ GeV) events for $p_T^{\gamma} \ga 20$ GeV, or even more for
lower cuts.
The $ZZH$ process is unfortunately an order of magnitude
smaller than the $WWH$.

\nn These cross sections are largest at a centre-of-mass
energy around $500$ GeV for $WWH$ and $ZZH$, almost independently of $m_H$
in the intermediate region. Although they fall off with energy the
decrease is not so drastic. For instance, at $1$ TeV they are about two times
smaller for $m_H \simeq 90$ GeV and are comparable for $m_H \simeq 150$ GeV.
This will be largely compensated for by the increase in integrated
luminosity,  since we contemplate at this energy a value
of about $60$ fb$^{-1}$. This means that at $\sqrt{s}=1$ TeV we will
 collect an even larger
 ``healthy" number of Higgs events than at $500$ GeV. For example, $HWW$
 will provide a welcome $200$ ( $m_H=90$ GeV) to
$130$ ($m_H=120$ GeV) additional Higgs events to those produced in the
conventional processes at 1 TeV.

\vskip 1.5cm
\noindent{ {\bf Acknowledgements} } \\
\noindent  We thank Guido Altarelli for useful discussions. Georges Azuelos
has been of invaluable help in promptly providing us with the {\em Pythia}
estimates of the $t \bar t$ background, including the cut on the $b \bar b$
invariant mass. We are very much indebted to
Edward Boos and Michael Dubinin for providing us with
the {\em ComHep} software.

\noindent
{\Large \bf Figure captions}\\

\noindent
\underline{Fig. 1} Feynman diagrams contributing to the processes
$e^+e^- \ra HW^+W^- \;(1a), HZZ \;(1b),\\
\noindent
HZ\gamma \;(1c)$ .\\
\noindent\underline{Fig. 2} Comparison among the different main
modes of single Higgs
production at a $500$ GeV $e^+e^-$ machine, as a function of the Higgs mass.
For $HZ\gamma$ we impose $p_T^\gamma>10$ GeV and $|y^\gamma|<2$.\\
\noindent
\underline{Fig. 3} Cross section for $e^+e^- \ra HW^+W^-$ as a function of the
centre-of-mass energy for three values of $m_H=90,120,150$ GeV (the
largest cross sections correspond to the smallest mass).\\
\underline{Fig. 4} As in the previous figure, but for $e^+e^- \ra HZZ$.\\
\noindent
\underline{Fig. 5} Cross section for $e^+e^- \ra HZ\gamma$ as a function of
the
centre-of-mass energy for the three values of $m_H=90,120,150$ GeV including
the cuts  $p_T^\gamma>10$ GeV and $|y^\gamma|<2$ (the
largest cross sections correspond to the smallest mass.)\\
\noindent
\underline{Fig. 6} $e^+e^- \ra HZ\gamma$ versus
the Higgs mass at $\sqrt{s}=300$ GeV and 500 GeV,
with the effect of different $p_T^\gamma$
cuts, but keeping $|y^\gamma|<2$. \\
\noindent
\underline{Fig. 7.1} Distribution in the transverse momentum of the photon in
the reaction $e^+e^- \ra HZ\gamma$ at $\sqrt{s}=500$ GeV,
for three values of the  Higgs mass. The cut $|y^\gamma|<2$ is applied.\\
\noindent \underline{Fig.7.2} Distribution in the energy of the $Z$ for
           $p_T^\gamma>10$ GeV in $e^+e^- \ra HZ\gamma$ at $\sqrt{s}=500$ GeV.
           The three curves correspond to three values of the Higgs mass as
           labelled in the previous graph.               \\
\noindent \underline{Fig. 7.3} As in Fig. 7.2, but with a cut
                $p_T^\gamma>30$ GeV. \\
\noindent \underline{Fig. 7.4} Distribution in the energy of the Higgs for
           $p_T^\gamma>10$ GeV in $e^+e^- \ra HZ\gamma$ at $\sqrt{s}=500$
GeV.\\
\noindent \underline{Fig. 7.5} As in Fig. 7.4, but with a cut
           $p_T^\gamma>30$ GeV. \\
\noindent \underline{Fig. 7.6} Distribution in the transverse momentum of
           the Higgs for
           $p_T^\gamma>10$ GeV in $e^+e^- \ra HZ\gamma$ at $\sqrt{s}=500$
GeV.\\
\noindent \underline{Fig. 7.7} As in Fig. 7.6,
but with a cut $p_T^\gamma>30$ GeV.\\
\noindent \underline{Fig. 7.8} Distribution in the angle between the beam and
            the direction of the Higgs for
           $p_T^\gamma>10$ GeV in $e^+e^- \ra HZ\gamma$ at $\sqrt{s}=500$
GeV.\\
\noindent \underline{Fig. 7.9} As in Fig. 7.8,
but with a cut $p_T^\gamma>30$ GeV.\\
\noindent \underline{Fig. 7.10} Distribution in the angle between the beam and
the
           direction of the $Z$ for
           $p_T^\gamma>10$ GeV in $e^+e^- \ra HZ\gamma$ at $\sqrt{s}=500$
GeV.\\
\noindent \underline{Fig. 7.11} As in Fig. 7.10,
but with a cut $p_T^\gamma>30$ GeV.\\
\noindent \underline{Fig. 7.12} Distribution in the angle between the $Z$ and
        the Higgs for $p_T^\gamma>10$ GeV in $e^+e^- \ra HZ\gamma$ at
          $\sqrt{s}=500$ GeV.\\
\noindent \underline{Fig. 7.13} As in Fig. 7.12,
but with a cut $p_T^\gamma>30$ GeV.\\
\noindent \underline{Fig. 7.14} Scatter plot in the energy of the Higgs $E_H$
    versus the energy of the photon for $p_T^\gamma>10$ GeV and
       $m_H=90$ GeV at $\sqrt{s}=500$ GeV.\\
\noindent \underline{Fig. 7.15} Scatter plot in the energy of the Higgs
    $E_H$ versus
the energy of the photon for $p_T^\gamma>10$ GeV and $m_H=90$ GeV at
                  $\sqrt{s}=1$ TeV.\\
\noindent \underline{Fig. 7.16} As previously, but with $m_H=150$ GeV and
                     $p_T^\gamma>30$GeV. \\
\noindent \underline{Fig. 8.1} Distribution in the transverse momentum of the
         Higgs in the process $e^+e^- \ra ZZH$ at
                  $\sqrt{s}=500$ GeV.\\
\noindent \underline{Fig. 8.2} Distribution in the angle between the beam and
the
              Higgs in the reaction
                $e^+e^- \ra ZZH$ at $\sqrt{s}=500$ GeV.\\
\noindent \underline{Fig. 8.3} As in Fig. 8.2
for the distribution in the angle
                 between
                the Higgs and the most energetic of the two $Z$. \\
\noindent \underline{Fig. 8.4} As in Fig. 8.3,
but with respect to the least energetic
                   $Z$.\\
\noindent \underline{Fig. 8.5} Scatter plot in the angle between the Higgs
               and the beam versus the angle between the most energetic $Z$
               and the beam in $e^+e^- \ra ZZH$ at $\sqrt{s}=500$ GeV for
               $m_H=90$ GeV. \\
\noindent \underline{Fig. 9.1} Distribution in the transverse momentum of the
         Higgs in the process $e^+e^- \ra W^+W^-H$ at $\sqrt{s}=500$ GeV.\\
\noindent \underline{Fig. 9.2} Distribution in the energy of the
         Higgs in the process $e^+e^- \ra W^+W^-H$ at $\sqrt{s}=500$ GeV. \\
\noindent \underline{Fig. 9.3} Distribution in the angle between the beam and
       the Higgs in the reaction $e^+e^- \ra W^+W^-H$ at $\sqrt{s}=500$ GeV.\\
\noindent \underline{Fig. 9.4} As in Fig. 9.3,
but concerning the angle between the
             electron and $W^-$.\\
\noindent \underline{Fig. 9.5} Scatter plot in  the $W^-$ energy
versus the angle between the
             electron and the $W^-$  for $m_H=90$ GeV
              in $e^+e^- \ra W^+W^-H$ at $\sqrt{s}=500$ GeV. \\
\noindent \underline{Fig. 9.6} Scatter plot in the angle between the beam and
            the Higgs versus the angle between the electron and the $W^-$
             for $m_H=90$ GeV in $e^+e^- \ra W^+W^-H$ at $\sqrt{s}=500$ GeV. \\

\end{document}